\documentclass{aa}  

\usepackage{graphicx}
\usepackage{txfonts}
\usepackage{amsmath}	
\usepackage{amssymb}	
\usepackage{xcolor}
\usepackage{hyperref}
\hypersetup{
    colorlinks=true,
    citecolor=blue,
    filecolor=blue,
    linkcolor=blue,
    urlcolor=blue
}
\newcommand{\dd}{\mathrm d}

\renewcommand{\(}{\left(}
\renewcommand{\)}{\right)}

\renewcommand{\v}[1]{\boldsymbol{#1}}

\newcommand{\Msun}{\,\mathrm{M}_\odot}

\begin{document}

   \title{How interacting winds shape the mechanical feedback of massive star clusters over millions of years}
   \titlerunning{Mechanical feedback of massive star clusters}
   \authorrunning{T. Vieu et al.}

   \author{T. Vieu\fnmsep\thanks{\email{thibault@mpi-hd.mpg.de}}, 
            L. H\"arer \fnmsep\thanks{\email{lucia.haerer@mpi-hd.mpg.de}}, and B. Reville}

   \institute{Max-Planck-Institut für Kernphysik, Saupfercheckweg 1, 69117 Heidelberg, Germany}

   \date{Received XXX XX, XXXX; accepted XXX XX, XXXX}

 
  \abstract
  {
   In recent years, massive star cluster environments have proved to be bright sources of very-high energy $\gamma$-rays, in particular young clusters which are powered by the winds interacting in their cores. 
   In order to understand how these winds can accelerate particles up to very-high energies, it is necessary to model their interactions from small (sub-pc) to large (10s of pc) scales over several millions of years. A key open question concerns the structure and properties of the resulting wind termination shock.
   By performing 3D magneto-hydrodynamic simulations of clustered winds embedded in a superbubble cavity,
   we demonstrate that the dynamics of stellar wind interactions and the resulting shock structure solely depends on the density and pressure of the cavity. This implies that the initial conditions of the simulation can be tuned in order to simulate star clusters of arbitrary age at a reduced computational cost. This novel method is validated using a toy cluster hosting 30 identical stars. We discuss the properties of the resulting cluster-wind termination shock under various assumptions. In particular, we are able for the first time to obtain a fully decoupled spherical wind termination shock for a 5 Myr old cluster. We further show that radiative cooling increases the sphericity of the shock. In general, the morphology of the outflow depends on the number of dominant stars, on the power of the stars sitting at the edge of the cluster core, and on the compactness of the cluster. We additionally show how a semi-analytical model can be used in order to estimate key morphological properties of the outflow without relying on large-scale simulations.
   {}
   }
   

   \maketitle
%

\section{Introduction}
\label{sec:intro}
Most massive stars spend their lives within stellar clusters, which are the main drivers of stellar feedback in the Galaxy. Powerful winds from young clusters plough through the interstellar matter, carving large-scale structures known as superbubbles \citep[e.g.][]{chu2008}, which constitute the hot phase of the interstellar medium (ISM). While the swept-up ISM accumulates at the edge of these cavities, the stellar winds interact in their depths, forming an ensemble of highly structured outflows and shock systems which, to date, remain poorly understood.

This powerful mechanical feedback has multiple consequences on the evolution of a galaxy. At large-scales, the expansion of superbubbles generates interstellar turbulence and creates over-densities in the ISM, which can subsequently trigger new episodes of star formation \citep{lee23}. If sufficiently powerful, superbubbles can break through the galactic disk, thereby forming direct advection routes feeding the halo \citep{kooMckee1992a}. At smaller-scales, \textcolor{black}{star cluster environments} accelerate particles, contributing to the non-thermal content of the Galaxy \textcolor{black}{and producing $\gamma$-rays via inverse-Compton scattering and proton-proton collisions. Indeed,} many massive star clusters have been observed to correlate with \textcolor{black}{diffuse} TeV $\gamma$-ray sources \citep[e.g.][]{hess11-wd2,gammaCygnus_HAWC2021,hess22-wd1, hess24-lmc,lhaaso24-w43,lhaaso24-cygnus}, \textcolor{black}{to the point that it has been argued that this novel class of sources could substantially contribute to the diffuse $\gamma$-ray flux \citep[e.g.][]{Menchiari2025} and could complement the picture of cosmic-ray production at isolated supernova remnants \citep[e.g.][]{Peron2024}. The mechanism by which particles are accelerated efficiently in star-cluster environments is however not yet understood. A promising scenario is that of particle acceleration at the cluster wind-termination shock, a large-scale shock surrounding the cluster that is predicted from a simple spherical-symmetry model (i.e. point-like injection of the cluster energy). In reality, energy injection by multiple stars is scattered over a significant volume and asymmetries are expected to be reflected in the structure of the cluster wind-termination shock. This is expected to have important consequences for particle acceleration \citep{Haerer2025}. For instance, it might happen that the sparse distribution of massive stars prevents the formation of a collective termination shock \citep{Vieu2024CygnusSimu}. The detailed structure of the cluster outflow surrounding the cluster core also becomes of crucial importance when it comes to the dynamics of supernova remnants expanding outwards from the core, one of the most promising scenarios to produce multi-PeV cosmic rays \citep{vieu2022Emax,Vieu2023,Anjos2025,Haerer2025Cygnus}.}

In a young massive star cluster, the mechanical energy is \textcolor{black}{in fact} dominated by only a few massive stars. These sources are separated by non-negligible distances (fractions of a parsec) and must therefore be considered discretely. Their energy is first thermalised in the cluster core through wind-wind interactions, then, provided the cluster core is compact enough, the flow is re-accelerated outwards to form a supersonic cluster outflow beyond the boundaries of the core. This collective outflow is shocked a few tens of parsecs away from the core, converting again the kinetic energy to internal pressure that inflates the superbubble cavity to hundreds of parsecs. Finally, the outflow is again re-accelerated at the edge of the superbubble, where it imparts direct mechanical feedback onto the ISM. Although these steps of energy reprocessing can be easily modelled in spherical symmetry \citep[e.g.][]{weaver1977}, they become non-trivial when an asymmetric cluster configuration, with discrete sources, is taken into account. Asymmetries are expected to decrease the efficiency of wind-wind interactions in the core, leading to incomplete mixing of stellar material and asymmetric expansion of the cluster outflow beyond the core \citep[e.g.][]{Scherer2018}. This can hinder the development of a spherical cluster wind termination shock, and modulate the spectrum of accelerated particles \citep[e.g.][]{Haerer2025} compared to predictions obtained in 1D \citep[e.g.][]{morlino2021}. Asymmetries over larger scales can also lead to energy losses, typically through enhanced radiative cooling near the edge of the superbubble, and thus decrease the overall efficiency of the mechanical feedback onto the ISM \citep[e.g.][]{gentry2019,lancaster2021}.

In order to quantitatively assess the mechanical feedback of massive star clusters onto their surrounding medium as well as their ability to accelerate non-thermal particles, we therefore require a detailed description of wind interactions, which are inherently three-dimensional. Analytic models \citep[e.g.][]{dyson72,weaver1977,chevalier1985,canto2000,Scherer2018,Owocki2025} as well as simulations modelling the star cluster as a homogeneous deposition of energy \citep[e.g.][]{krause2013,Rogers2013,gupta18, elbadry2019,lancaster2021} are unable to fully capture the shock dynamics or to predict the properties of the magnetic fields. Only high-resolution simulations, with injection of individual winds as kinetic sources at sub-pc scales, can attempt to describe wind-wind interactions and their consequences at large-scale. This has so far only been discussed in \citet{Raga2001,RodriguezGonzalez2007,badmaev2022,badmaev23,Vieu2024core}, who restricted the simulation to the cluster core over 10s kyr; \citet{Haerer2025}, who described the structure of the wind termination shocks after several 100s kyr, and \citet{Vieu2024CygnusSimu} who reached Myr timescale but in the special case of the loose association Cygnus OB2. We are therefore still lacking a simulation showing the development of cluster-wind termination shocks over several Myrs, which corresponds to the age of the most powerful clusters seen in TeV $\gamma$-rays.

The present work aims at bridging the timescale gap using a novel initial condition for 3D MHD simulations of stellar clusters. After describing the simulation setup (Section~\ref{sec:setup}), we perform a first, ``traditional'' simulation of a star cluster in a homogeneous medium (Section~\ref{sec:HDSimuNascent}). We then describe our novel initial condition that allows us to skip the first few Myr of evolution (Section~\ref{sec:superbubble_ansatz}). This method is validated and used to simulate clusters in various configurations and up to an age of 10\,Myr. In Section~\ref{sec:analyticstuff} we discuss analytical characterisations of the three-dimensional asymmetries in the cluster wind. We conclude in Section~\ref{sec:conclusions}.

\section{Simulation setup}
\label{sec:setup}
In order to solve the dynamics of wind-wind interactions in a compact massive star cluster, we perform large-scale MHD simulations. The MHD equations are solved using the publicly available PLUTO code \citep{PLUTO2007}. Technicalities of the setup (solver, time-stepping, divergence cleaning) were already detailed in Section~2.1 of \citet{Haerer2025}. In the following we only highlight the changes compared to this previous work.

\subsection{Star cluster}
\label{sec:setup_cluster}
In order to simplify the parameter space and showcase the key mechanisms of stellar-wind interactions without ad-hoc assumptions, we implement in the present work a toy cluster of 30 identical stars homogeneously distributed in a core of radius 2.5~pc (for the base run). All stars have mass-loss rate $\dot{M} = 3 \times 10^{-6}\,\Msun$/yr, terminal wind-velocity $\varv_{\infty} = 2500$~km/s, wind sound speed $c_s = 23$~km/s, surface magnetic field of 100\,G, rotational velocity at the equator of 300\,km/s and a stellar radius of 20\,R$_\odot$. This represents a total mechanical power of $1.78 \times 10^{38}$\,erg/s, close to values typical of young massive star cluster \citep[e.g.][]{parizot2004}.

The orientation of the rotation axis as well as the positions of the stars in the cluster core are randomly generated. For reproducibility purposes, the list of positions and orientations is provided in Appendix~\ref{sec:appendixclustersetup}.

\subsection{Stellar-wind injection}
Individual stellar winds are injected by forcing the velocity, density and pressure at the boundary of spheres with a 5 cell radius:
\begin{align}
    \v{u} = \varv_{\infty} \v{e_{r,i}} \, , \qquad 
    \rho = \frac{\dot{M}}{4 {\rm \pi} r_i^2 \varv_{\infty} } \, , \qquad 
    P = \rho c_s^2 \, ,
\label{injectionprofile}
\end{align}
where $\v{e_{r,i}}$ is the radial unit vector of the spherical coordinate system centred on star $i$ and $r_i$ the distance from the star $i$.
The magnetic field is prescribed with a Parker spiral \citep[][see Eqns. (8) and (9) in \citealt{Haerer2025}]{parker58}. At run initialisation we further define a transition zone beyond the injection boundary of 5 cells up to a radius of 10 cells. In this transition zone all MHD variables are connected linearly to their environmental values. As remarked in \citet{Haerer2025}, this is necessary to avoid the expansion of ``boxy'' stellar winds when used with Cartesian grids. \citet{Haerer2025} used a different, physically-motivated prescription in the transition region, however further testing showed that implementing linear profiles is sufficient, with the advantage that it removes initial discontinuities.

Note that Equation~\ref{injectionprofile} is only valid if the wind's ram pressure dominates the thermal pressure at the boundary of the injection region. The grid resolution in the core region must be fine enough to ensure that this criterion is always fulfilled.

\subsection{Grid}
\begin{table*}
	\centering
\begin{tabular}{l|l|l|l}
Core radius [pc]  & Side-length of central region [pc]  & Resolution in central region [pc/cell] & Resolution at box edge [pc/cell]   \\
		\hline
		\hline
1.25  & 4.5 & 0.025 & 6.30 \\
2.5 (default) & 9.0 & 0.05  & 5.43 \\
5.0 & 9.0 & 0.05 & 5.43  \\
10.0  & 18.0 & 0.1 & 4.29  \\
	\end{tabular}
	\caption{Box edge resolutions are given for the 5 Myr old cluster (box size 120\,pc).}
	\label{tab:run_resolutions}
\end{table*}

As in \citet{Haerer2025} we use a non-uniform grid with a homogeneous central region and logarithmic spacing beyond. 
The grid is stretched but not nested, which has the advantage to preserve single cell interfaces, with the drawback that the cells become strongly elongated along the Cartesian axes. This can introduce numerical artifacts at large distances. It is however not expected to impact the dynamics of wind-wind interactions close to the cluster core. 

The resolution in the central region depends on the adopted size of the core, see Table~\ref{tab:run_resolutions}. The total size of the box is set such that the entire wind-blown superbubble is simulated: 65\,pc for 1\,Myr old cluster, 120\,pc for the 5\,Myr old cluster, and 170\,pc for the 10\,Myr old cluster.

\subsection{Radiative cooling and heating}
\label{sec:setupcooling}
Optically thin cooling and heating are implemented as reaction terms in the equation for the internal energy:
\begin{equation}
   \frac{\partial e}{\partial t} = ... - n_H^2 \Lambda(T) + n_H n_0 X_H \Lambda(T_0) \, , 
\end{equation}
where $n_0$ and $T_0$ are respectively the initial (ISM) density and temperature, $n_H$ is the hydrogen number density of the plasma and $\Lambda$ the cooling function.
We assume that the plasma is fully ionised (mean molecular weight $\mu = 0.61$) in the entire simulation box (including the ISM) with an hydrogen mass fraction $X_H = 0.711$. We adopt the default tabulated cooling function generated in PLUTO, \textcolor{black}{ignoring the ionisation state of the gas and the chemical reaction network, as our aim is not to accurately capture the detailed evolution of the large-scale superbubble shell, but to qualitatively model the collapse of the shell, and its destabilisation leading to enhance mass-loading at the interface \citep{elbadry2019}.}

The heating term is designed to maintain the ISM at the equilibrium temperature $T_0=5 \times 10^3$~K. \textcolor{black}{We assume a constant heating rate per particle in the entire simulation box, agnostic to the heating process.}
Due to numerical errors it may happen that cells with very low pressure appear close to strong shocks. These are only a few cells which have no impact on the simulation result, however they are problematic as their associated heating rate is very large, which dramatically slows down the overall computation. In order to alleviate this numerical issue, we cap the heating rate when the heating time falls below 1\,yr.

Thermal conduction is not included as solving a parabolic equation on top of the MHD closure is too computationally intensive. As discussed in Appendix~\ref{sec:appendix_tc}, tests performed over very restricted timescales ($< 50$\,kyr) show that adding thermal conduction would not dramatically modify the dynamics of the outflows \citep[see also sec. 3.2 of][]{badmaev2022}. The impact of thermal conduction over longer timescales is however most likely non-negligible and should be quantitatively assessed in future works, especially if one aims to predict the superbubble density, temperature, and shell instabilities at large distances from the cluster core. This requires dedicated simulations beyond the scope of the present work.

\section{Simulation of a nascent star cluster up to 1 Myr}
\label{sec:HDSimuNascent}

\begin{figure*}
    \centering
  	\includegraphics[width=\linewidth]{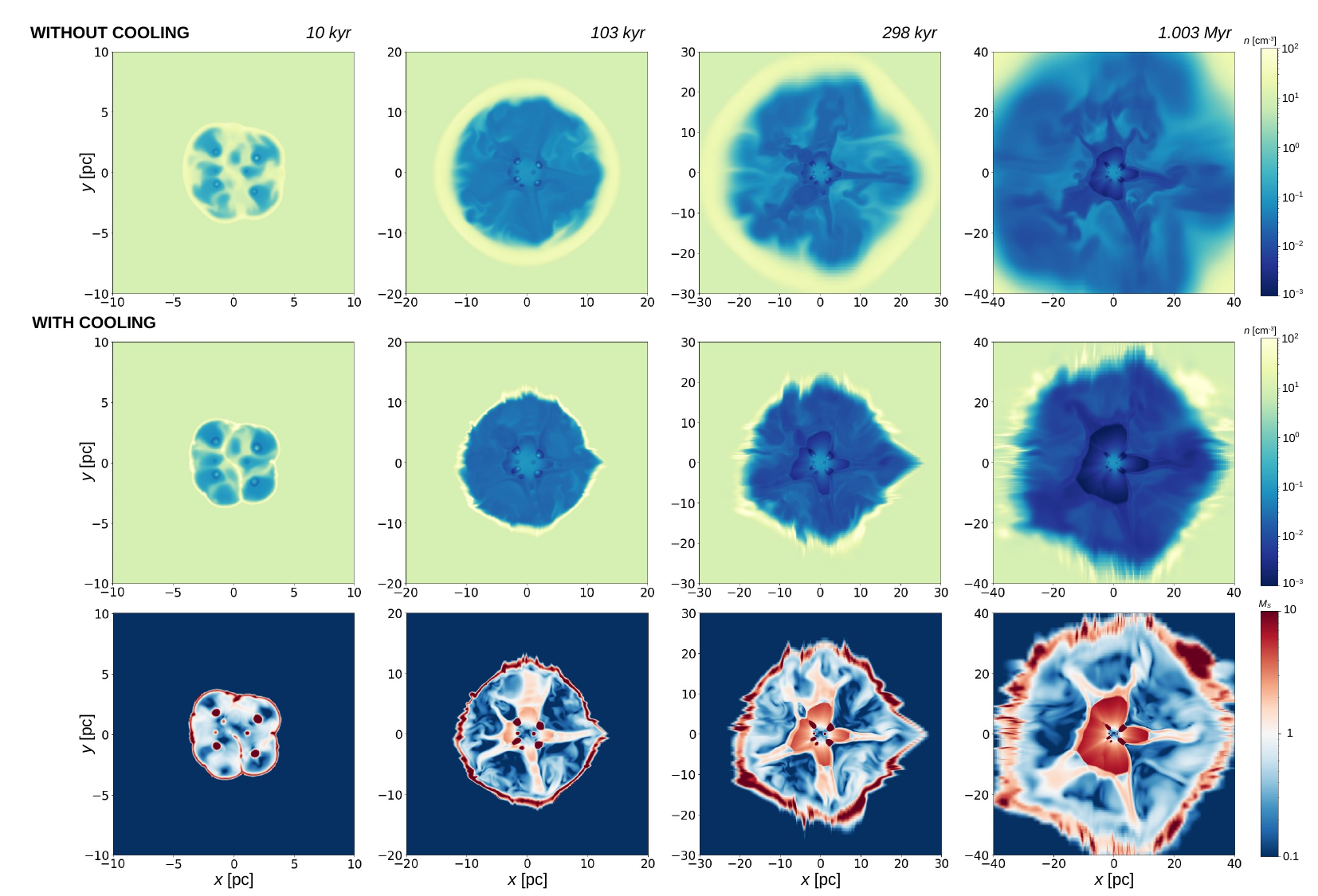}
  	\caption{Evolution of the stellar feedback over 1 Myr in a homogeneous ISM. First row: density slices without cooling. Second row: density slices with cooling. Third row: Mach number with cooling. Timestamps from left to right: 10kyr, 103 kyr, 298 kyr, 1.003 Myr. \textcolor{black}{The striations aligned with the Cartesian axes visible at 1\,Myr are numerical artifacts due to the stretching of the grid.}}
  	\label{fig:slice_evolutionfromt0}
\end{figure*}

We first perform a control simulation by injecting stellar winds in a homogeneous ISM of density $n_0=10$\,cm$^{-3}$. \textcolor{black}{This is an idealisation as young star clusters are expected to evolve into turbulent clumpy HII regions shaped by early UV radiation feedback, photoevaporation and mechanical feedback \citep[e.g.][]{Mellema2006,Kim2018}. As a result, our simulation is not expected to provide a detailed description of the evolution of the superbubble at large scales ($\sim 100$\,pc). Rather, our aim is to probe shock structures at small and intermediate scales ($\sim 10$\,pc) in the hot superbubble interior which, independently of the external conditions, is expected to be homogeneous to a good approximation.}

\textcolor{black}{We make another simplification by implementing a toy cluster of coeval stars, as mentioned in Section~\ref{sec:setup_cluster}, ignoring stellar evolution, such that the power output is constant in time.} 

\textcolor{black}{From these initial conditions, we let the system evolve up to an age of 1\,Myr.}
Stellar interactions quickly create a cavity with a complex inner structure whose evolution is shown in Figure~\ref{fig:slice_evolutionfromt0}.

\subsection{Transsonic sheets and cluster termination front}

\begin{figure*}
    \centering
  	\includegraphics[width=0.8\linewidth]{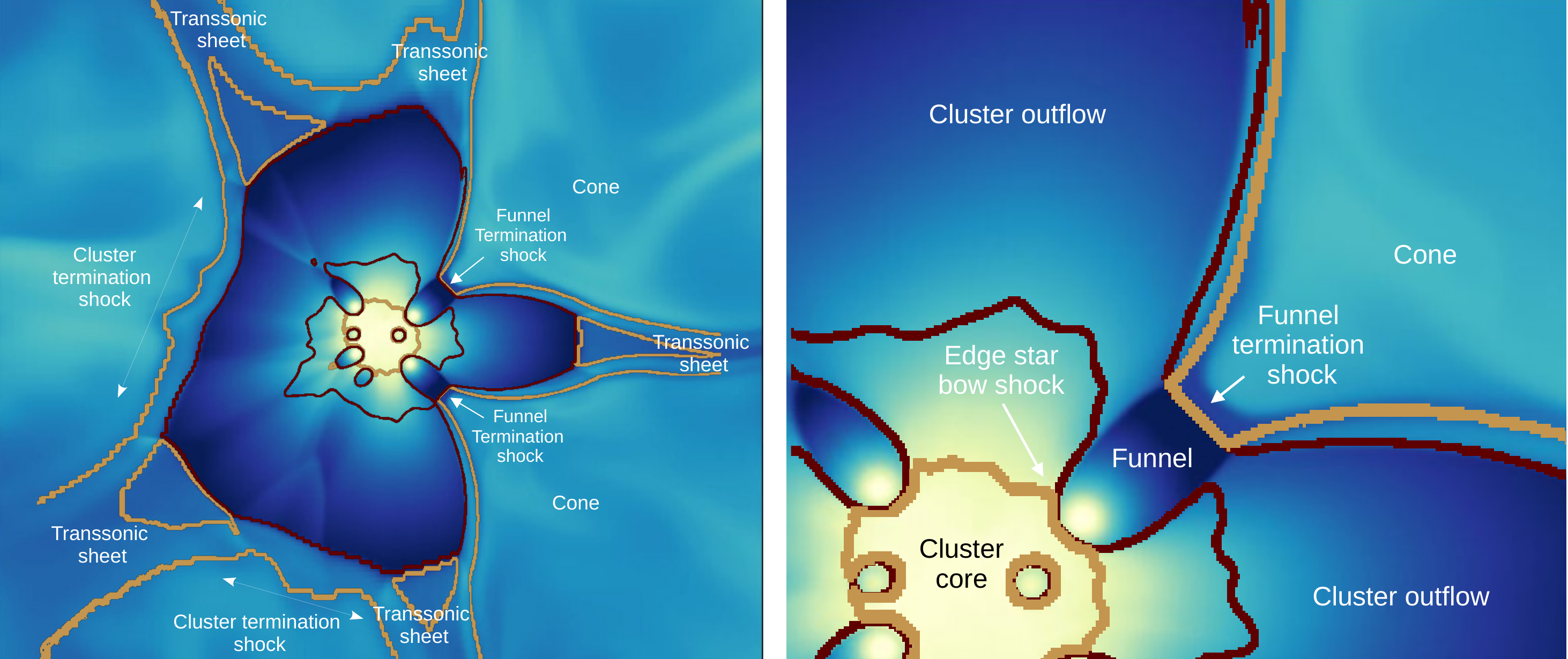}
  	\caption{Detailed structure of the cluster outflow and cluster termination front at 1 Myr. Strong supersonic contours (Mach number = 3) are shown in red. Transsonic contours (Mach number = 1) are shown in orange.}
  	\label{fig:WTS_scheme_full}
\end{figure*}

Individual cavities are first excavated around each star. They quickly expand, collide and merge into a collective superbubble. Remaining density inhomogeneities dissipate within 10\,kyr, leaving the star cluster at the heart of a low-density environment bounded by a dense spherical shell. In the cluster core, wind-wind collisions generate a stationary turbulent bulk flow as described in dedicated simulations \citep{badmaev2022,Vieu2024core}. The bulk flow is subsonic within the core but re-accelerates outwards due to the pressure gradient between the core (high pressure) and the superbubble (lower pressure). Contrary to one-dimensional predictions, this outward expansion does not produce a spherical flow beyond the cluster core, as the stellar winds lying on the cluster edge act as obstacles which mechanically obstruct the flow. The latter is deflected and channelled between the edge stars. As a result, two-dimensional trans-sonic sheets are launched in the superbubble.
A strong shock is present at the base of these sheets (see Mach number slices in Figure~\ref{fig:slice_evolutionfromt0}). This starts to resemble the ``cluster wind termination shock'' expected from one-dimensional theory \citep[e.g.][]{dyson72,weaver1977,chevalier1985,canto2000}. However the geometry of the strong shock is far from spherically symmetric, especially at early times (a few 100\,kyrs). As the cluster evolves, the strong shock expands further away from the core and the thermal pressure of the collective wind eventually dominates over the ram pressure of the individual winds' shocks, which form droplet shapes that eventually ``detach'' from the cluster shock, a process that we call ``decoupling''. As this happens, the termination front of the cluster outflow (``cluster termination front'' in short) gets closer to spherical symmetry, although the shock surface remains very inhomogeneous at 1\,Myr.

Figure~\ref{fig:WTS_scheme_full} details the structure of the cluster outflow at 1\,Myr. To the left of the slice, all edge stars have already been overtaken by the collective outflow, i.e. their individual wind termination shocks have decoupled. The collective outflow terminates here at the cluster termination shock. This shock is only moderately strong ($M_s \lesssim 10$) as there is a strong oblique velocity component in the direction of the trans-sonic sheets. Downstream of the cluster termination shock, the outflow is completely subsonic and mixes with the hot superbubble material. This is different within the trans-sonic sheets, where the collimated flow is re-accelerated in a narrow low density channel and forms a succession of Mach surfaces similar to those seen in supersonic-aircraft engine exhausts (see the Mach number plot at 103\,kyr in Figure~\ref{fig:slice_evolutionfromt0}). Finally, the two edge stars on the right of the core are still far from being decoupled. Their winds funnel directly through the collective outflow and thermalise only when they collide with the hot superbubble medium, forming a small, but very strong, ``funnel termination shock''. In this region, the global shock surface strongly bends inwards in a conical shape.

We note that such a complex flow structure is to be expected when discrete sources are considered, i.e. when the mechanical power is dominated by a small number of stars. Incomplete mixing in the core and asymmetric outflow channels are purely kinetic effects which were already described in detail in \citet{Scherer2018}. Wind-wind collisions create a network of separatrix surfaces which prevent mixing. Several edge-star winds fail to thermalise in the bulk outflow, but instead directly collide with the superbubble medium.
This is in stark contrast to the spherically symmetric picture often assumed. Ultimately, such models are based on the assumption of complete thermalisation of the individual winds in the cluster core, and full mixing of the wind material \citep{canto2000,Owocki2025}. These assumptions are however not justified. 

We do not repeat here the discussion on the properties of the wind termination shock (magnetic field, obliquity, Mach number...) that was conducted in \citet{Haerer2025}. Instead, we focus on two additional effects that impact the geometry of the shock, namely radiative cooling and late time evolution, as both were not covered in that work.

\subsection{Effect of cooling}
The middle row of Figure~\ref{fig:slice_evolutionfromt0} shows the same simulation described above and shown in the first row, but now with radiative cooling enabled (see Section~\ref{sec:setupcooling} for implementation details). The qualitative evolution is similar, with notable differences near the superbubble shell.

As expected from theory, the superbubble shell quickly collapses onto a thin layer once the simulation time reaches the shell cooling time ($t_c \sim 13 \left( L_{38}^3 / n_0^8 \right)^{1/11}$\,kyr, \citealt{kooMckee1992a}). The shell is too unstable to retain a perfectly spherical shape and is prone to be deformed by the trans-sonic sheets launched from the cluster core. In our simulation, instabilities arise mainly due to the grid stretching at large radii (this produces artifacts visible in the shell, see Figure~\ref{fig:slice_evolutionfromt0}). In reality one expects a number of sub-grid processes to make the shell corrugated or even break it. In any case, this induces additional mixing of material from the shell into the superbubble, as proposed in \citet{elbadry2019} \citep[see also][for complementary 3D simulations]{lancaster2021,lancaster2024}.

\begin{figure}
    \centering
  	\includegraphics[width=\linewidth]{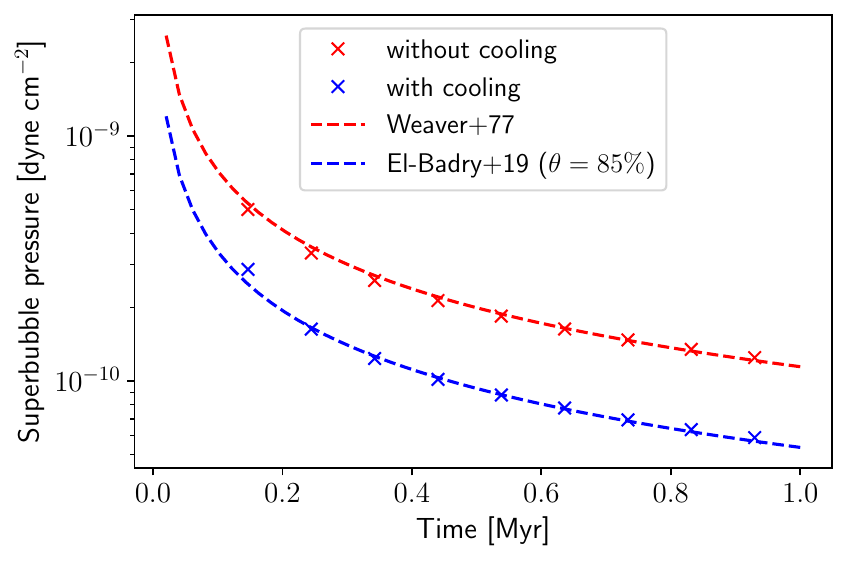}
  	\caption{Evolution of the mean pressure in the subsonic superbubble medium.}
  \label{fig:pressure_evolution_intime}
\end{figure}

This additional mixing happens over a wide transition layer close to the inner interface of the shell. This enhances cooling of the superbubble interior. As a result, the pressure inside the superbubble is decreased by a factor of about 2, the size of the superbubble is significantly smaller than expected from analytic predictions and the cluster wind termination shock can expand up to larger distances.
This agrees quantitatively with the model described in \citet{elbadry2019}, which leads to the following modifications of the superbubble radius and pressure:
\begin{align}
    R_{\rm SB} &\approx 80 \( \frac{(1-\theta) L_c}{10^{38} {\rm erg/s}} \)^{1/5} \( \frac{n_0}{{\rm cm}^{-3}} \)^{-1/5} \( \frac{t}{{\rm Myr}} \)^{3/5} \,\rm{pc} \label{radiusSBwithcooling}
    \\
    P_{\rm SB} &\approx 1.7 \times 10^{-11} \( \frac{(1-\theta) L_c}{10^{38} {\rm erg/s}} \)^{2/5} \( \frac{n_0}{{\rm cm}^{-3}} \)^{3/5} \( \frac{t}{{\rm Myr}} \)^{-4/5} \,\rm{dyne/cm}^2 \label{pressureSBwithcooling}
\end{align}
As shown in Figure~\ref{fig:pressure_evolution_intime}, an excellent fit to our simulation data is obtained assuming $\theta=0.85$, i.e. 85\% of the total cluster power is lost in radiative cooling.

Although the parameter $\theta$ is expected to depend on sub-grid shell instabilities that cannot be resolved in our simulation, the observable result on the superbubble size and the morphology of the wind termination shock should be weakly affected as the superbubble pressure scales in theory as $((1-\theta) L_c)^{2/5}$. In conclusion, radiative cooling enhanced by turbulent mixing in the vicinity of the superbubble shell facilitates the expansion of the cluster-wind termination shock by lowering the internal pressure. Because the superbubble is depressurised more rapidly, it becomes pressure-confined in the external medium at an earlier stage.

\textcolor{black}{Although enhanced mass-loading from mixing at the shell interface is an expected consequence of radiative cooling, we point out that our simulation setup is not optimised to quantitatively study the large-scale evolution of the superbubble. More predictive simulations on the topic have been recently performed by \citet{lancaster2024,lancaster2025}.}

\section{Simulating evolved star clusters from a superbubble ansatz}
\label{sec:superbubble_ansatz}

\begin{figure*}
    \centering
  	\includegraphics[width=\linewidth]{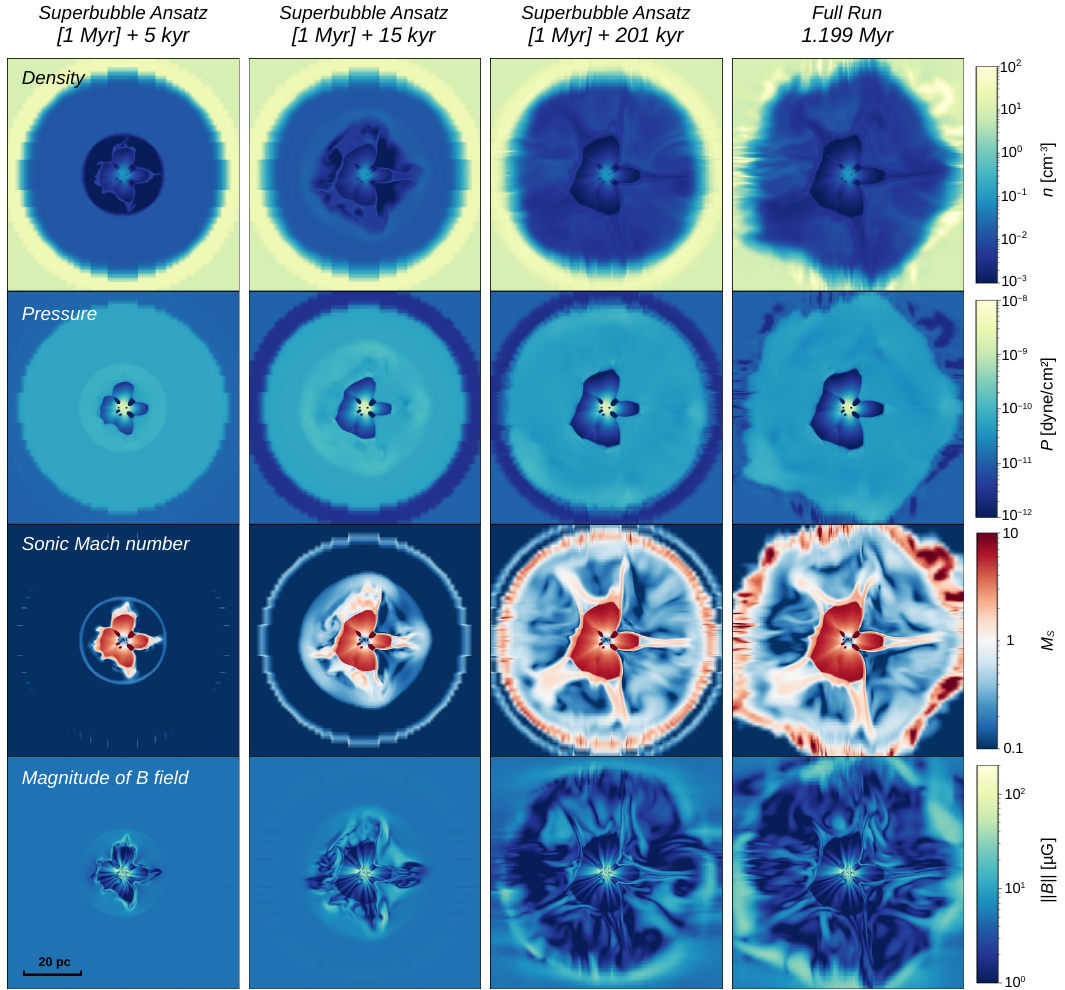}
  	\caption{Evolution of the stellar feedback over 200 kyr starting with a superbubble ansatz (3 first columns), compared with the solution obtained in the full run starting at $t=0$ (last column).}
  	\label{fig:WeaverAnsatzProofat1Myr}
\end{figure*}

Simulating a star cluster starting from $t=0$ is expensive. Each run described in the previous section took around 25 000 CPU-hours (about 48h on 512 processors) on the MPCDF-HPC Raven, which is already long for a rather low core resolution. Simulating a more compact cluster (necessitating higher core resolution) up to later times (e.g. 5 Myr) would require about 10 times more resources, which becomes less feasible, especially if one wishes to explore the parameter space (e.g. vary the compactness of the cluster core, the number of stars etc.).

On the other hand, our simulations are geared towards investigating the structures arising from wind interactions deep inside the superbubble in cases where the cluster configuration deviates from a point-like injection of energy. The expansion of the cluster termination front is solely driven by the interaction between stellar winds (which is well captured in our setup) and the pressure of the superbubble medium against which the cluster termination front stalls. 
Independently of the complexity of the superbubble solution at large-scale, the superbubble interior must be isobaric since the sound crossing time in this hot medium is smaller than the dynamical time: any perturbation is expected to disappear on $\sim 100$s\,kyr timescales. Therefore the morphology of the wind termination shock of a given cluster solely depends on the \textit{average} superbubble pressure, while the detailed structure of the large-scale superbubble (in particular its shell dynamics) is not directly relevant. It is worth noting in particular that the only reason the cluster-wind termination front expands in time is because the superbubble pressure decreases over time.

We should therefore be able to reduce our set of initial assumptions to i) a cluster configuration and ii) a value for the superbubble pressure. This is the main idea of the ``superbubble ansatz'' method described below.

\subsection{Initialisation of a superbubble ansatz}
Instead of modelling an external medium as the initial condition of the simulation, we now prescribe a ``superbubble ansatz'' as follows. In addition to the cluster power $L_c$ and total mass loss rate $\dot{M}_c$, we define the outer superbubble radius, $R_{\rm SB}$, the radius of the shell interface, $R_{\rm CD}$, and the pressure inside the superbubble, $P_{\rm SB}$. These three parameters can be either guessed from analytic estimates [e.g. from \citealt{weaver1977}, possibly correcting for cooling as $L_c \to (1-\theta) L_c$], or spherical simulations. They can also be taken as input parameters since $R_{\rm CD}$ and $R_{\rm SB}$ ultimately encode the shell dynamics and ISM properties, and $P_{\rm SB}$, which results from the cumulative input of stellar winds throughout the entire cluster's history, can be interpreted as a redefinition of the cluster age. Note that in the simplest case of a steady cluster power and homogeneous ISM, the equivalence between pressure and time is given by Equation~\ref{pressureSBwithcooling}. In the following we will present results for given ``equivalent times'' with the notation $t=[t_0] + t_1$. The time $t_0$ within brackets is the age of the superbubble at run start, obtained using Equation~\ref{pressureSBwithcooling} from the assumed pressure $P_{\rm SB}$. The time $t_1$ outside of the brackets is the time which has been actually simulated. For instance, [5 Myr] + 0.5 Myr would mean that the superbubble radius and internal pressure were initialised assuming a cluster age of 5\,Myr, and that the system was then simulated during 0.5\,Myr, starting from the ansatz. The result is then expected to approximately match that of a full computation over 5.5\,Myr.

Once the superbubble parameters have been prescribed, the ansatz is implemented as follows.
In a spherical region of radius $R_{\rm SB}$ around the centre of the domain, the pressure is set to the uniform value $P_{\rm SB} > P_0$, where $P_0$ is the average pressure of the external medium (e.g. $P_0 = 7 \times 10^{-12}$\,dyne/cm$^2$ for $T_0 = 5 \times 10^3$\,K and $n_0 = 10$\,cm$^{-3}$). To initialise the density field, we further define 3 zones: the inner region for $r<R_{\rm WTS} = \sqrt{ \dot{M}_c \sqrt{2 L_c/\dot{M}_c} /(4 {\rm \pi} P_{\rm SB}) }$, the superbubble for $R_{\rm WTS}<r<R_{\rm CD}$ and the shell for $R_{\rm CD}<r<R_{\rm SB}$. The density inside the shell is obtained from conservation of the swept-up mass: $n = n_0/\left[1-(R_{\rm CD}/R_{\rm SB})^3\right]$. The average density inside the superbubble is obtained from conservation of the mass launched by the stellar winds over the entire cluster history: $\rho_{\rm SB} \approx \int \dd t \dot{M}_c /(4{\rm \pi} R_c^3/3)$. The density in the inner region is taken as a wind profile: $\rho = 0.0563 \dot{M}_c^{3/2} L_c^{-0.5}/R_{\rm WTS}^2 $. The velocity field is set to zero everywhere in order to avoid the creation of voids at the beginning of the simulation. While this initial condition is not fully consistent with the continuity equation, the wind output from the cluster will quickly sweep-up the initial density gradient and converge towards a physical solution. Setting up an initial density gradient in the inner region facilitates the early expansion of the wind and speeds up the convergence to the quasi-stationary solution.

\subsection{Proof-of-principle}
\label{sec:proofprinciple}
\begin{figure*}
    \centering
  	\includegraphics[width=0.8\linewidth]{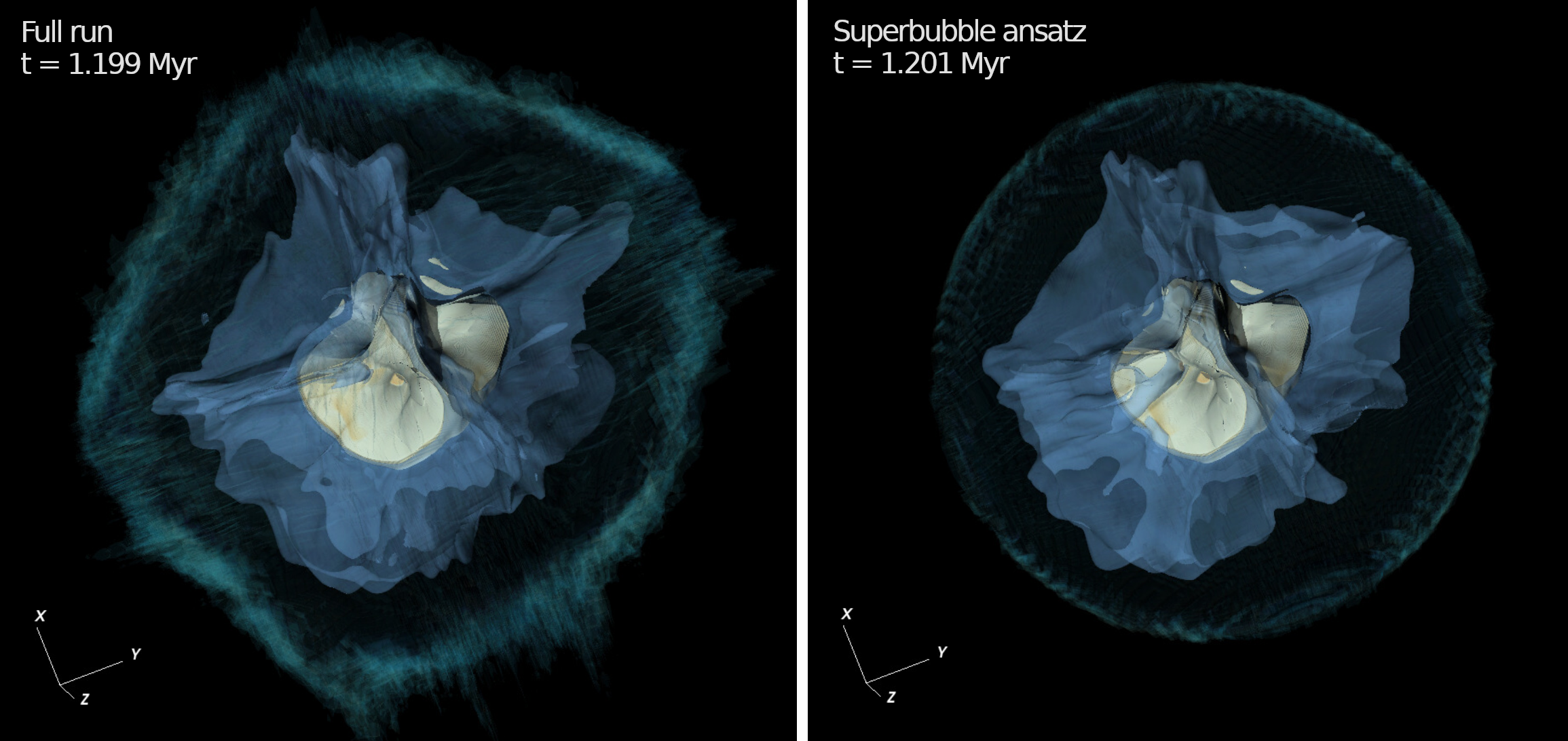}
  	\caption{Comparison between the solution obtained at 1.2 Myr in the full run and the solution obtained at 200 kyr in the run started with the 1 Myr old superbubble ansatz. The white surface is the cluster termination front. The blue sheets are the trans-sonic sheets escaping the cluster core. The turquoise interface renders the outer shell of the superbubble.}
  	\label{fig:3Dcomparison1Myr}
\end{figure*}

Figure~\ref{fig:WeaverAnsatzProofat1Myr} shows how the simulation evolves when the cluster is initialised using the superbubble ansatz. The pressure in the superbubble is set to $6 \times 10^{-11}$\,dyne/cm$^2$, the inner interface of the superbubble shell is placed at $R_{CD} = 36$\,pc and the outer shell is placed at $R_{SB} = 42$\,pc. These parameters are tuned to the average values obtained in the simulation of Section~\ref{sec:HDSimuNascent} after 1~Myr of evolution. As seen in Figure~\ref{fig:WeaverAnsatzProofat1Myr}, the stellar winds quickly expand in the low-density inner region. A pressure wave is launched in the superbubble which reaches the shell within one sound crossing time and bounces back toward the centre, while the cluster termination front settles to its quasi-stationary structure. After a couple of sound crossing times, the simulation has converged to a quasi-stationary state which closely resembles the output of the full simulation started at $t=0$ (see right panels in Figure~\ref{fig:WeaverAnsatzProofat1Myr} and Figure~\ref{fig:WeaverAnsatzProofat1Myr_otherslices} for the other slices, as well as the 3D view comparison in Figure~\ref{fig:3Dcomparison1Myr}). In particular, the shape of the cluster termination front is well reproduced. The magnetic field structure is also well reproduced in the inner region, and displays qualitatively similar large scale features in the bubble, with similar intensities. A full agreement in the bubble is not expected as we initialised an unmagnetised superbubble ansatz.

This demonstrates two things. First, it is possible to simulate star cluster termination shocks starting from an arbitrary superbubble age, provided one can estimate the corresponding superbubble pressure. This alleviates the difficulty of running a full simulation up to a given age. Not only does this speed up the computation by a factor of about 10, but also, as discussed earlier, an assumption on the superbubble pressure is more robust than an assumption on the age. It alleviates the necessity to make ad-hoc assumptions on the evolution of the cluster and the environment. On the other hand, it is always possible to write an equivalence between pressure and age assuming an evolution model.

Second, our result demonstrates that the past evolution of a cluster has no impact on the shape of the cluster termination front provided the cluster has been in a slowly-evolving state over the last few 100~kyrs. ``Slowly-evolving'' here means that any dynamical process happens over a time-scale longer than one or two sound crossing times from the shock to the superbubble edge. In particular, stellar proper motion of a few km/s is not expected to impact the structure of the collective termination shock. Past Wolf-Rayet \textcolor{black}{or} supernova activity will only affect the average superbubble pressure at a given time, not the details of the wind structure (unless a recent supernova exploded, in which case the cluster termination front is expected to be disturbed).

In conclusion, the structure of a cluster termination front at an arbitrary age solely depends on the average pressure in its surrounding superbubble and any change of cluster configuration quickly relaxes to a quasi-stationary solution that doesn't depend on the detailed evolution of the system from $t=0$.

\subsection{Simulation of evolved stellar clusters}
\begin{figure}
    \centering
  	\includegraphics[width=\linewidth]{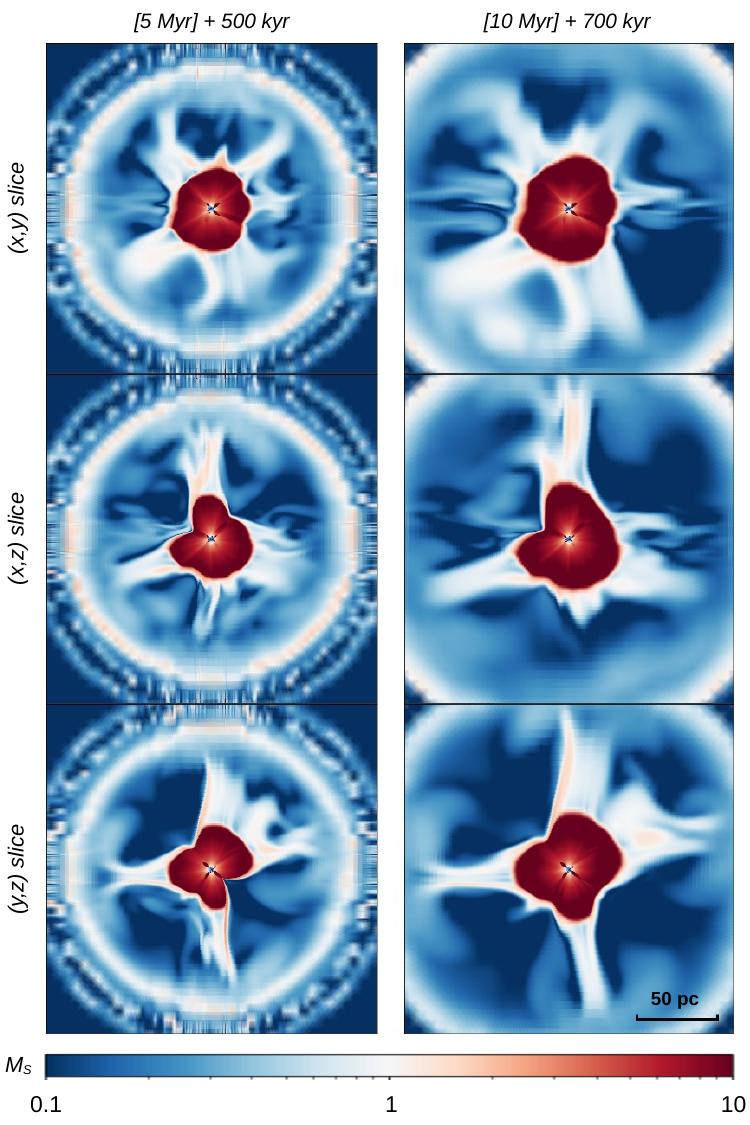}
  	\caption{Sonic Mach number solutions obtained starting with a 5 Myr old superbubble ansatz (left) and with a 10 Myr old superbubble ansatz (right). The simulations have converged after 0.5 Myr (left) and 0.7 Myr (right). The three rows show different slices. Note that the figure bounding box is kept the same for all slices: the right panels are not a zoom of the left panels but a simulation of an older cluster.}
  	\label{fig:comparison_VaryAge}
\end{figure}

The superbubble ansatz method allows us to simulate stellar clusters of arbitrary age (i.e. evolving in superbubbles of arbitrary pressure). We perform additional simulations assuming the average superbubble pressure and radius follow the modified Weaver solution (Eqs~\ref{radiusSBwithcooling}, \ref{pressureSBwithcooling}). We initialise two clusters, at 5 and 10 Myr old respectively. After initialising the superbubble ansatz, we let the interacting winds evolve until the system reaches a quasi-stationary state. This takes typically one sound-crossing time, that is, 0.5\,Myr for the 5\,Myr old cluster and 0.7\,Myr for the 10\,Myr old cluster. The result is shown in Figure~\ref{fig:comparison_VaryAge}.

\begin{figure*}
    \centering
  	\includegraphics[width=\linewidth]{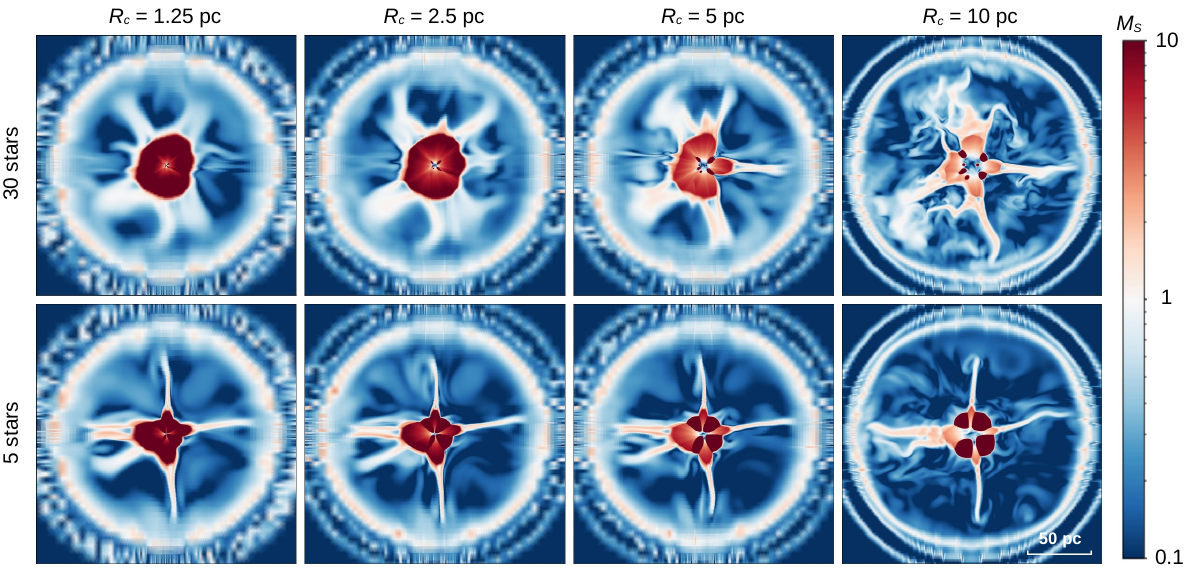}
  	\caption{Solutions obtained for a [5\,Myr]+500\,kyr old cluster, varying the  compactness (left to right: 1.25, 2.5, 5 and 10 pc) and number of dominant stars (30 for the top panels and 5 for the bottom panels), keeping the total power the same. Upon rescaling the core, the relative positions of the stars are kept the same.}
  	\label{fig:comparison_VaryCompact}
\end{figure*}

\begin{figure*}
    \centering
  	\includegraphics[width=\linewidth]{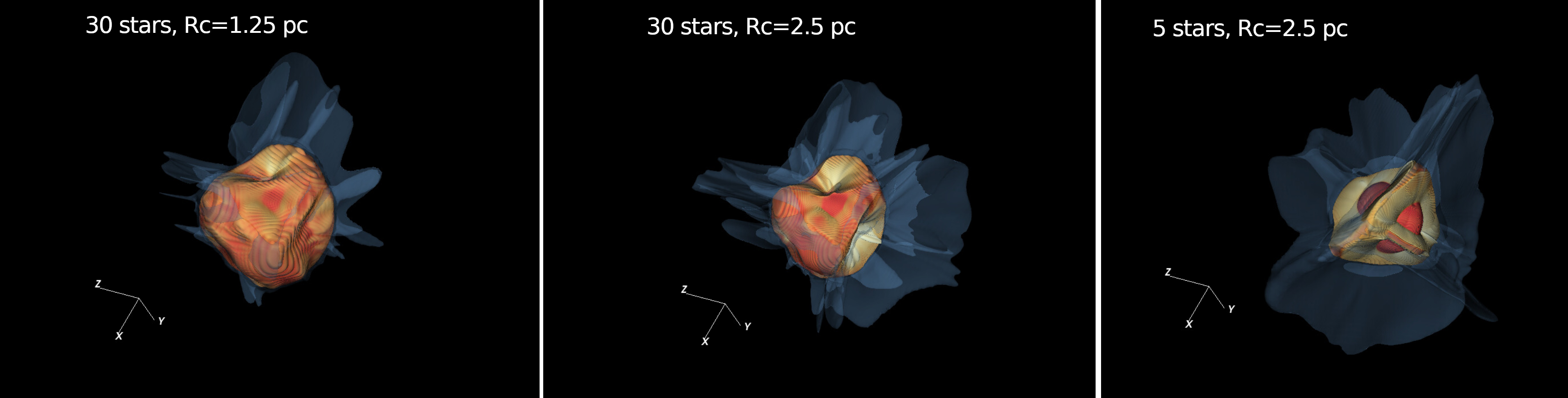}
  	\caption{3D rendering of the solutions for the most compact and homogeneous cluster (left), the nominal cluster (centre), and a cluster dominated by only 5 stars (typically 5 toy Wolf-Rayet stars, right). All clusters are [5\,Myr]+500\,kyr old.}
  	\label{fig:3Dview5Myr}
\end{figure*}

At 5\,Myr, the cluster termination front is almost fully decoupled from individual winds, and at 10\,Myr, it is fully decoupled. This cluster termination front is significantly more spherical than for the 1\,Myr old cluster.
In order to characterise the shock geometry in a quantitative way, we define two proxies: the decoupling fraction and the coefficient of inhomogeneity. The decoupling fraction is the ratio of the shock surface crossed by the cluster collective flow to the total shock surface:
\begin{equation}
    DF = 1 - S_{\rm individual}/S_{\rm CWTS} \, ,
\end{equation}
where $S_{\rm individual}$ is the surface of the shock crossed by the flow coming from a single star and $S_{\rm CWTS}$ is the total surface of the cluster wind termination shock.

The coefficient of inhomogeneity quantifies the shock sphericity:
\begin{equation}
    CI = 2 \sqrt{\langle R(\theta,\phi)^2 \rangle}/\langle R(\theta,\phi) \rangle \, ,
\end{equation}
where $R(\theta,\phi)$ is the shock surface. As a rule of thumb, a coefficient of inhomogeneity below 10\% means that the shock is close to spherical symmetry, a coefficient of inhomogeneity above 30\% means that the shock is significantly aspherical, and a coefficient of inhomogeneity above 50\% means that the shock is extremely aspherical.

For the 5\,Myr old cluster with core radius 2.5\,pc, we obtain $CI = 34\%$, i.e. the shock is still significantly aspherical. For the 10\,Myr old cluster with core radius 2.5\,pc, we obtain $CI = 28\%$. This demonstrates how difficult it is to produce a spherical cluster termination front even when its mean radius is much larger than the extent of the cluster core. 

\subsection{Impact of cluster compactness and number of stars}
\label{sec:compactnessandnumberofstars}

In Figure~\ref{fig:comparison_VaryCompact} we compare the results of eight simulations, varying the radius of the cluster core from 1.25\,pc to 10\,pc, and the number of stars (30 or 5, keeping the same total cluster power). Although a star cluster hosting only 5 stars might seem to represent an edge case, it is in fact not implausible that a handful of stars dominate the power output of a massive star cluster. Indeed, a typical massive star cluster of age between 5 and 10\,Myr is expected to host a few powerful Wolf-Rayet stars. In fact, the nominal cluster of 30 stars might be more of an edge case, representative only of the most massive star clusters, for example Westerlund 1 (24 Wolf-Rayet stars).

\begin{figure}
    \centering
  	\includegraphics[width=\linewidth]{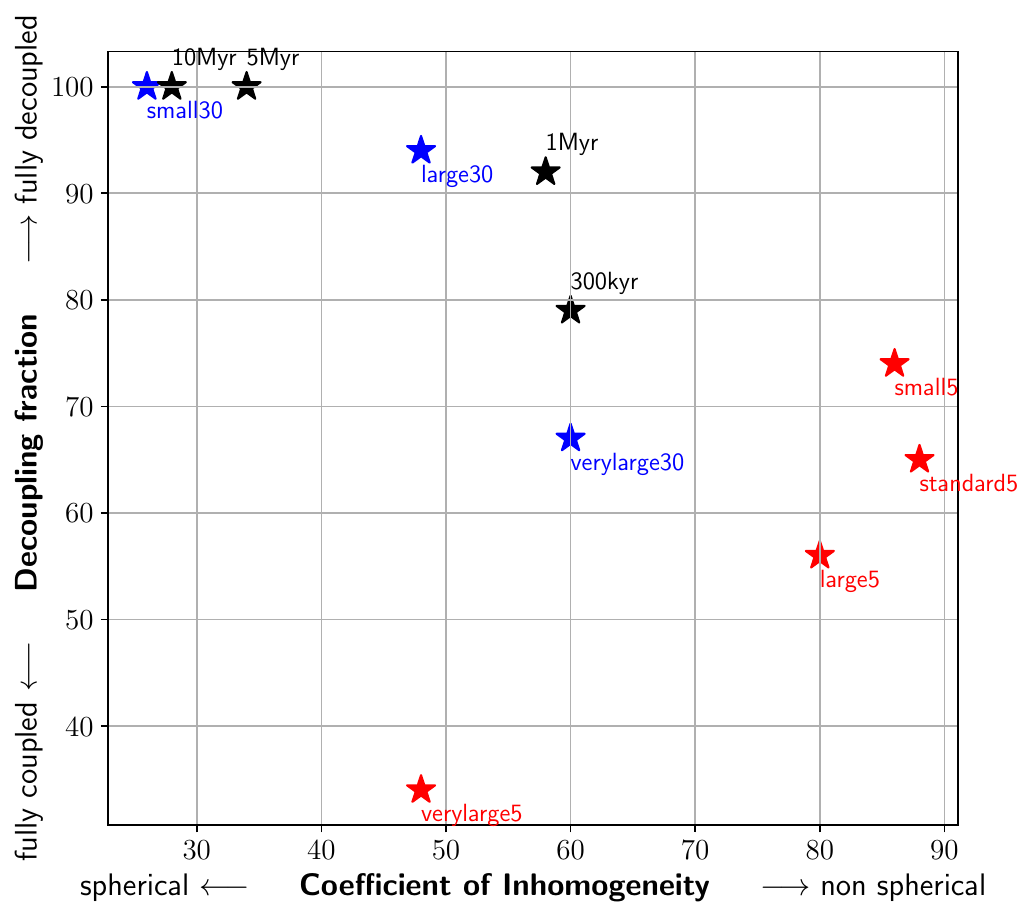}
  	\caption{Properties of the cluster termination surface for all simulated clusters. As a rule of thumb , a coefficient of inhomogeneity above 30 already implies significant asymmetry. Black markers show the standard cluster with 30 stars in a core radius of 2.5\,pc, taken at different ages. Blue (resp. red) markers show additional simulations for 5\,Myr old clusters hosting 30 (resp. 5) stars. Size tags refer to the core radius: 1.25\,pc for ``small'', 2.5\,pc for ``standard'', 5\,pc for ``large'' and 10\,pc for ``very large''.}
  	\label{fig:diagram}
\end{figure}

It is expected that the larger the cluster core, the longer it takes for the edge stars to decouple. At the simulated age of 5\,Myr, only the clusters with core radii 1.25 and 2.5\, pc are fully decoupled (see also the 3D views displayed in Figure~\ref{fig:3Dview5Myr}). For each simulation, we calculated the decoupling fraction and the coefficient of inhomogeneity. The result is summarized by the diagram in Figure~\ref{fig:diagram}. Only the smaller cluster is reasonably spherical. Clusters of 5 stars are far both from being decoupled and from spherical symmetry.
 
\begin{figure}
    \centering
  	\includegraphics[width=\linewidth]{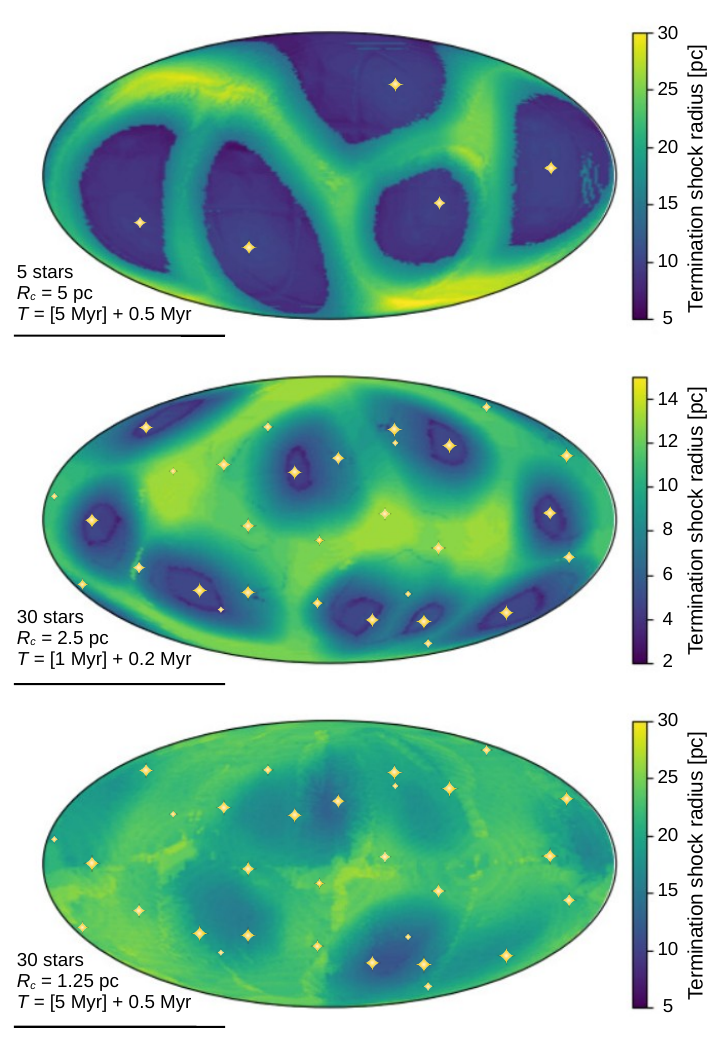}
  	\caption{Mollweide projection showing the radius of the wind-termination shock along line outs. The positions of the stars are shown by the markers, whose sizes scale with the distance of the stars to the center of the cluster (i.e. big markers are edge stars). Top: large cluster core radius, 5 Myr 5 stars. Middle: standard 1 Myr. Bottom: Small 5 Myr 30 stars. }
  	\label{fig:skymaps}
\end{figure}

We also show sky maps in Figure~\ref{fig:skymaps} which display the radius of the termination shock as a function of the spherical angles in a Mollweide projection. Dark blue regions correspond to individual wind termination shocks still coupled to the cluster termination shock. Note that the coefficient of inhomogeneity defined earlier is simply a measure of the inhomogeneity of these maps. It is clear that in the case of a large core of 5\,pc with only 5 dominant stars, individual wind-termination shocks still make up for a significant fraction of the cluster termination front at 5\,Myr. For the nominal run ($2.5$\,pc core for 30 stars at 1\,Myr), we identify 9 edge stars which are still coupled. Only by decreasing the size of the core and increasing the age can we obtain a fully decoupled cluster termination shock (bottom panel).

There is a striking similarity between the results obtained for various core radii at 5\,Myr, and the results previously obtained at various ages for a core radius of 2.5\,pc. For instance, the run with $R_c=5$\,pc at [5\,Myr]+500\,kyr (see Figure~\ref{fig:comparison_VaryCompact}) resembles the result obtained with $R_c=2.5$\,pc at [1\,Myr]+200\,kyr (see Figure~\ref{fig:WeaverAnsatzProofat1Myr}). The run with $R_c=10$\,pc at [5\,Myr]+500\,kyr (see Figure~\ref{fig:comparison_VaryCompact})  resembles the result obtained with $R_c=2.5$\,pc at 0.2\,Myr (see Figure~\ref{fig:slice_evolutionfromt0}). This is reminiscent of the self-similarity of the Weaver solution, where the radius of the spherical cluster wind termination shock (from a single star) scales as $t^{2/5}$.  

\section{Derivation of the decoupling time}
\label{sec:analyticstuff}
We observe in the simulations that individual winds from powerful stars located at the edge of the cluster core can remain coupled over a long time, in particular if the cluster is not compact or if the cluster power is dominated by a small number of stars. Complete decoupling is a necessary condition to obtain a truly collective cluster termination shock, close to spherical symmetry. In this section we therefore seek a criterion to estimate the decoupling time of a given cluster without having to run full hydrodynamic simulations.

\begin{figure}
    \centering
  	\includegraphics[width=\linewidth]{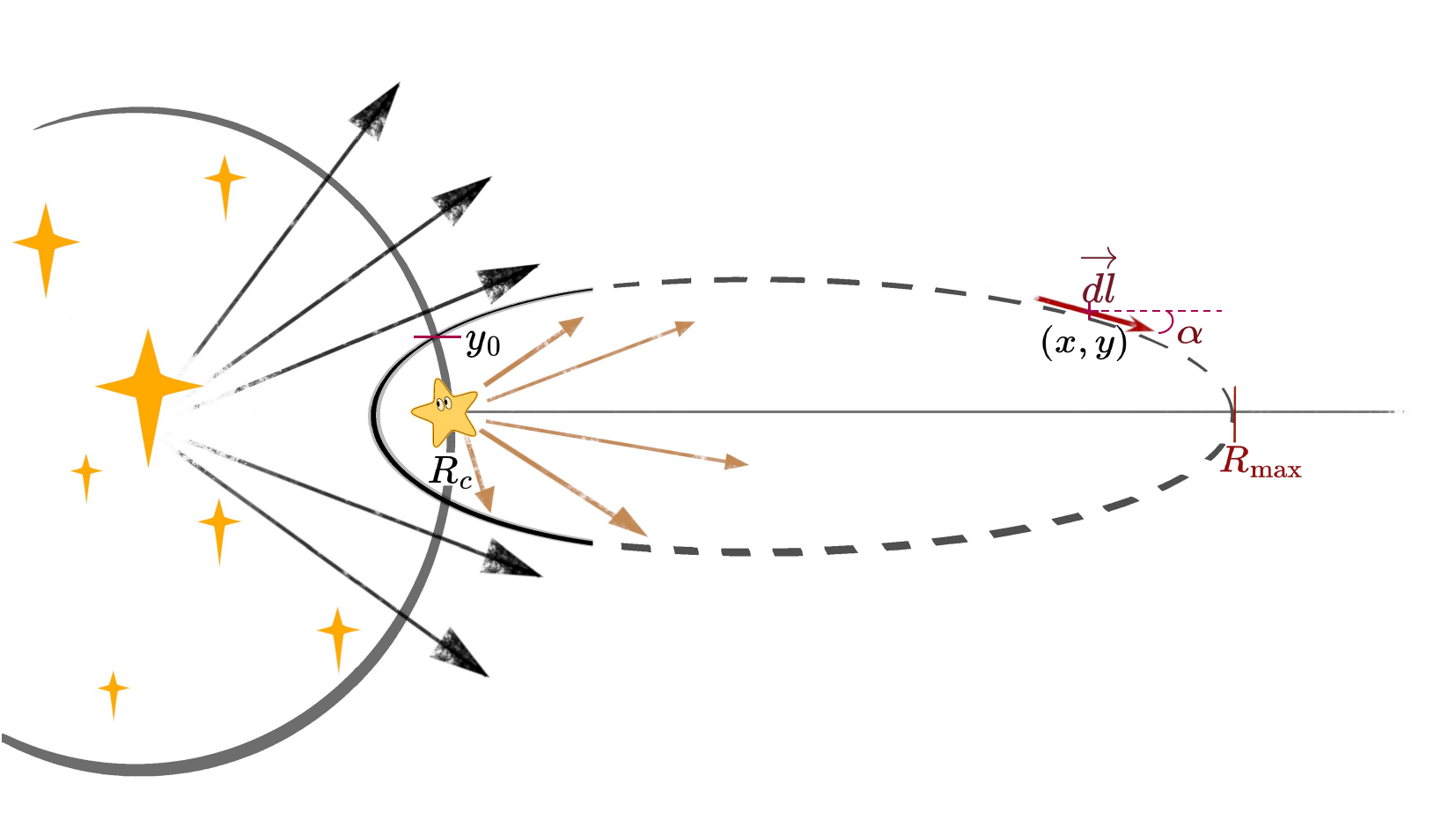}
  	\caption{Model for the funnelled wind behind a star positioned at the edge of the cluster core.}
  	\label{fig:decoupling_scheme}
\end{figure}

Let us consider a star located at the edge of the cluster core (see Figure~\ref{fig:decoupling_scheme}) and an infinitesimal segment of the interface between the funnelled wind of the edge star and the collective cluster wind at the position $(x,y)$. The segment is tangent to the unit vector $\v{dl} = (\cos \alpha, \sin \alpha)$. The orientation of the interface is determined by the balance between the ram pressure of the funnelled wind, the thermal pressure of the surrounding cluster wind, and possibly the ram pressure of the cluster wind if $\alpha>\theta = \arctan(y/x)$.

\citet{chevalier1985} have obtained an analytic expression for the thermal pressure in the collective wind beyond the cluster core in the case of a homogeneous deposition of energy in a spherical core. We assume in the following that this formula holds everywhere but in the funnelled wind of the edge star. This is a reasonable assumption if most of the cluster kinetic energy is deposited and thermalised within the core (i.e. there must be enough powerful stars positioned deep in the core), and if the decoupling fraction is already large. The latter condition is realised by definition when we approach the decoupling time.

The ram pressure of the edge star wind against the interface reads:
\begin{equation}
    P_{*}^{\rm ram} = \rho_* u_*^2 \frac{\( (R_c-x) \sin \alpha + y \cos \alpha \)^2}{(x-R_c)^2 + y^2} \, .
\end{equation}
The ram pressure of the cluster wind against the interface reads:
\begin{equation}
    P_c^{\rm ram} = \rho_c u_c^2 \sin^2(\theta-\alpha) \Theta(\alpha-\theta)  \, ,
\end{equation}
where $\Theta$ is the Heaviside function.
The thermal pressure in the cluster wind reads, for $r>R_c$:
\begin{equation} \label{chevaliercleggpressure}
    P^{\rm th}_c = \chi(r/R_c) \dot{M_c}^{1/2} L_c^{1/2} R_c^{-2} \, , \quad \chi(r) = A r^{-10/3}  \, ,
\end{equation}
where $\rho_c = \dot{M_c}/(4 {\rm \pi} u_c r^2)$ is the density in the cluster wind, $\rho_* = \dot{M_*}/(4 {\rm \pi} u_c ((x-R_c)^2 + y^2))$ is the density in the edge star wind, $\dot{M_c}$ is the cluster mass-loss, $\dot{M}_*$ is the edge star mass loss, $u_c$ is the cluster wind velocity, $u_*$ is the edge star wind velocity. We also defined the radius $r=\sqrt{x^2+y^2}$ and polar angle $\theta = \arctan(y/x)$. Equation~\ref{chevaliercleggpressure} comes from \citet{chevalier1985} where $A = 0.0106$ is a numerical factor.

Working out the equilibrium condition between the three terms leads to:
\begin{equation*}
    \frac{\sin^2(\theta-\alpha) \Theta(\alpha-\theta)}{R^2} + 
    \frac{4 {\rm \pi}}{\sqrt{2}} 
    \chi(R) = \eta \left( \frac{(1-X) \sin \alpha + Y \cos \alpha}{(1-X)^2 + Y^2} \right)^2
\end{equation*}
where $\eta \equiv \sqrt{ \dot{M}_* L_*/ \dot{M}_c L_c }$ is approximately the power ratio between the edge star and the cluster. We have also introduced the dimensionless variables $X=x/R_c$, $Y=y/R_c$, $R=r/R_c$. 

The initial position of the interface is at $x\approx R_c$, and $y$ such that the ram pressure of the edge star is equal to the thermal pressure at the edge of the core. The later reads $P^{\rm th}_c(r=R_c) = \xi \dot{M_c}^{1/2} L_c^{1/2} R_c^{-2}$ with $\xi = 0.0338$ \citep{chevalier1985}. This provides the starting coordinate:
 \begin{equation}
     Y_0 = \frac{\sqrt{2} \eta}{4 {\rm \pi} \xi}  \, .
 \end{equation}
 Taking this as our initial condition we can numerically integrate the following differential equation:
 \begin{equation}
     \frac{\dd \v{X_I}}{\dd s} = \v{dl} = (\cos \alpha, \sin \alpha)  \, ,
 \end{equation}
 to find the curve $\v{X_I}$ that delimits the interface between the funnelled wind and the cluster wind. In particular we obtain $R_{\rm max}$ such that the interface reaches the axis $y=0$ for $X = R_{\rm max}$. Note that the only physical parameter that appears in the above equations is $\eta$, which implies that the geometry of the interface is invariant under rescaling of the core.

 \begin{figure*}
    \centering
  	\includegraphics[width=\linewidth]{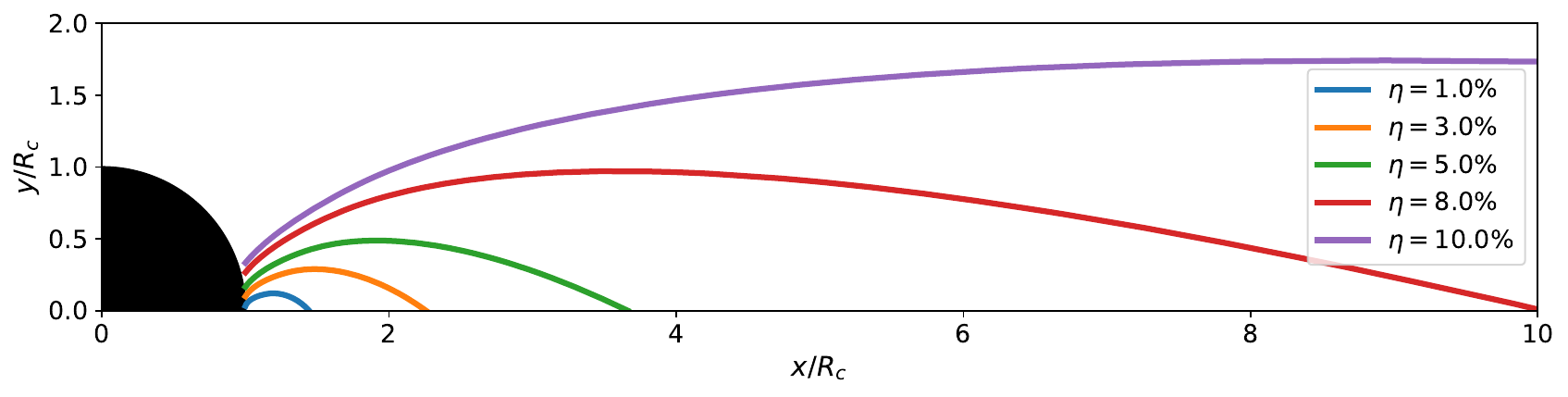}
  	\caption{Analytic profiles for the funneled wind behind an edge star. The curves trace the interface between the wind of the single star and the collective cluster wind. The thermal pressure of the latter bends the wind in the wake of the single star.}
  	\label{fig:wake_profiles}
\end{figure*}

Figure~\ref{fig:wake_profiles} shows the interface profile for various values of $\eta$. It can be shown that the profiles will always eventually fall back onto the y-axis, however this can happen at large distances from the core if the power of the edge star is a sizeable fraction of the cluster power.

 As the cluster wind evolves in time, the collective cluster wind as well as the funnel from the edge star progressively expand in the superbubble medium. Because the ram pressure of the cluster wind is larger than the ram pressure of the funnel outflow, the cluster wind termination shock expands more rapidly, hence creating the ``cones'' seen in the simulation behind the individual wind shocks (see Figure~\ref{fig:WTS_scheme_full}). The expansion of the funnel is driven by the balance between the ram pressure at the funnel termination shock, $P_{*}^{\rm ram} = \dot{M}_* u_*/(4 {\rm \pi} (x - R_c)^2$ and the thermal pressure in the superbubble, $P_{\rm SB}$. The decoupling therefore occurs when:
 \begin{equation}\label{decouplingtime}
     \frac{4 {\rm \pi} R_c^2}{\dot{M}_c u_c} P_{\rm SB} = \frac{\eta}{(R_{\rm max} - 1)^2} \, .
 \end{equation}
Finally the superbubble pressure is related to the age as $P_{SB} \propto t^{-4/5}$ (Eq.~\ref{pressureSBwithcooling}). As noted earlier, $R_{\rm max}$ is independent of $R_c$, therefore we have shown that the decoupling time scales as $R_c^{5/2}$. This is precisely the reason why we previously noted similarities between simulation results. The shape of the wind termination shock is primarily determined by the level of coupling of edge stars funnels, such that the shape obtained at a given time $t$, for a cluster core of size $R_c$, is similar to the shape obtained at a time $a^{5/2} t$, for a cluster core of size $a R_c$. For instance, doubling the core radius multiplies by 5.6 the decoupling time: this is why the run at 5.5\,Myr for a core of 5\,pc is similar to the run at 1\,Myr for a core of 2.5\,pc.

 \begin{figure}
    \centering
  	\includegraphics[width=\linewidth]{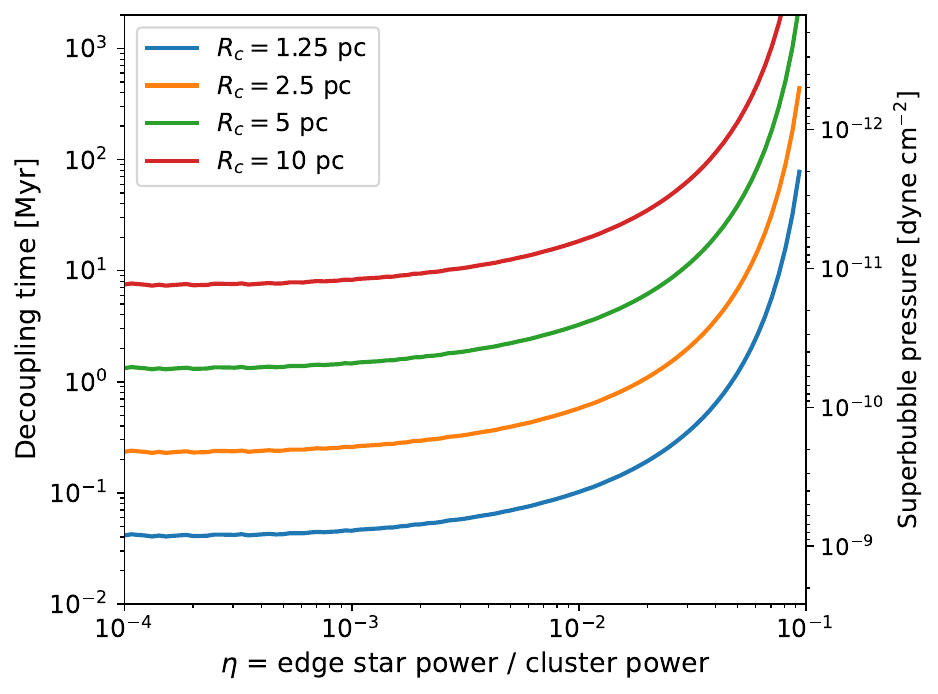}
  	\caption{Decoupling time of an edge star predicted by the analytic model.}
  	\label{fig:decouplingtimes}
\end{figure}

We can solve Equation~\ref{decouplingtime} assuming that the pressure in the superbubble follows Equation~\ref{pressureSBwithcooling} with $\eta=0.85$. The result is plotted in Figure~\ref{fig:decouplingtimes}. For the cluster with 30 stars ($\eta=1/30$), the theoretical decoupling times are 0.5\,Myr ($R_c=1.25$\,pc), 2.8\,Myr ($R_c=2.5$\,pc), 15.8\,Myr ($R_c=5$\,pc) and 89\,Myr ($R_c=10$\,pc). This agrees with the few constraints given by the simulations.

It should be noted that the pressure evolution law (Eq.~\ref{pressureSBwithcooling}) does not account for the pressure of the external medium. Once the superbubble has expanded to the point where the interior pressure equals the ISM pressure, the bubble stalls and the pressure does not further decrease. In fact, in our simulation setup, this should happen at about 10\,Myr. Therefore, in reality, the clusters with $R_c=5$\,pc or $R_c=10$\,pc are never expected to decouple.

Finally, the cluster with only 5 stars is never expected to decouple within any reasonable time. This challenges spherical models of cluster wind termination shocks, in particular in the context of particle acceleration \citep[e.g.][]{morlino2021}. If a few very massive stars, such as early O or WR stars, are located near the edge of the cluster core, a substantial fraction of the cluster termination front will be coupled to the funnel termination shocks behind these edge stars, and the truly collective wind solution will never develop within a reasonable time.

\bigbreak
The quantitative predictions of the above model are limited, as we used the \citet{chevalier1985} solution outside of its strict domain of validity, we neglected turbulence in the cluster wind, we assumed that edge stars are placed exactly at the core boundary, we assumed that the superbubble pressure is completely homogeneous even in the conical structures close to the funnel termination shocks, etc. Nevertheless we believe that this analysis provides reasonable order-of-magnitude benchmarks to determine if an edge star funnel is expected to decouple or not. In principle, this can also be applied to the case of a more realistic initial mass function. It enables reconstruction of the entire wind termination shock in 3D at a given age by calculating the position of the cluster-wind termination shock, the position of the funnels' termination shocks, and the orientation of the funnels' interfaces which, when prolonged beyond the funnels' termination shocks, trace the subsonic cones. 

\section{Conclusions}
\label{sec:conclusions}
Resolving asymmetric wind-wind interactions in the core of a compact massive star cluster over timescales of several Myrs is computationally challenging. We have shown that an approximate solution to the problem can be found by tweaking the initial conditions of the simulation, under the assumption that the winds efficiently clear out a hot ($\gg 10^6$\,K) cavity around the cluster. In such a hot medium, the sound speed is so high that, to a good approximation, the thermal pressure is homogeneous. On the other hand, the expansion of the supersonic cluster outflow is solely hindered by the superbubble pressure. Adjusting the initial pressure in the simulation therefore enables to simulate the collective wind of a star cluster which has already evolved up to an arbitrary age.

Importantly, this also shows that the geometry of the cluster outflow at a given age does not depend on the detailed evolution and dynamics of the cluster, only on the superbubble pressure at the considered age and the current stellar power. \textcolor{black}{More generally, one of the strengths of the superbubble ansatz method is that it is agnostic to the details of the superbubble evolution and its large-scale geometry. Because of the complexity of the system, the computational cost and the uncertainty on many parameters such as the evolution of young massive stars, it is futile to try including all relevant physical processes all at once over Myr timescales in a single high-resolution simulation. The ansatz method allows instead to reduce most of this complexity to a single number: the superbubble pressure at a given age. Then the main parameter of the system, and the most relevant, becomes the superbubble pressure, instead of the age.
Detailed simulations including additional physics such as thermal conduction, photoionization, or ISM turbulence, are needed to find sensible prescriptions for the superbubble pressure at a given age. However these simulations do not need to simulate individual stellar winds, so they can in principle afford a much lower resolution.}

Our novel simulation method allowed us to probe interactions of clustered winds up to an equivalent age of 10\,Myr, that is, a factor 10 to 100 compared to previous works. For the first time, we were able to go beyond the ``decoupling time'', that is, the time at which the collective cluster outflow overcomes the ram pressure of the individual edge stars such that a spherical cluster wind termination shock starts to expand. A reasonably spherical large-scale shock was however only obtained in the idealised case of a cluster hosting 30 homogeneously distributed identical stars within a compact core radius (at most a few parsecs). More extended clusters ($R_c \gtrsim 3$\,pc) and/or clusters hosting fewer powerful stars (e.g. only a handful of Wolf-Rayet stars) are never expected to decouple. In this case, mixing of wind material is inefficient, a spherical wind termination shock does not develop, and the shock surface is mostly made of conical structures bending inward, separated by two-dimensional sheets of transsonic flows.

These findings modify our understanding of shocks in the vicinity of compact star clusters and therefore question the validity of spherical models for the acceleration of particles in these environments. \textcolor{black}{The superbubble ansatz method will enable the possibility to perform multiple simulations in order to explore the parameter space at realistic cluster ages with the aim to investigate the shock asymmetry, the topology of the magnetic field and their consequences on particle confinement, non-thermal spectra and maximum reachable energies. This simulation method will also allow to study the detailed dynamics of supernova remnants expanding in the vicinity of compact clusters, which is key to validate recent analytical models claiming that such supernova remnants would be able to accelerate particle beyond PeV energies \citep{Vieu2023}.}

Finally we note that the superbubble ansatz method is limited by the assumption that the cavity carved by the cluster is reasonably spherical up to several tens of parsecs. Even though this is a standard assumption in the field, it might not be satisfied if the cluster lies within a dense and turbulent molecular cloud \citep[e.g.][]{Rogers2013,lancaster2021}, or a region with a large-scale density gradient (e.g. if the star cluster is significantly offset from the galactic plane).

\begin{acknowledgements}
      This work made use of the MHD code PLUTO. Computations were performed on the HPC system Raven at the Max Planck Computing and Data Facility.
\end{acknowledgements}

%
%

\bibliography{biblio}
\bibliographystyle{aa}


\begin{appendix}
\section{Toy cluster}
\label{sec:appendixclustersetup}
Table~\ref{tab:starlist} provides the list of stars implemented in the simulation. All stars have identical mass-loss rate $\dot{M} = 3 \times 10^{-6}\,\Msun$/yr, wind terminal velocity $\varv_{\infty} = 2500$~km/s, wind sound speed $c_s = 23$~km/s, surface magnetic field of 100\,G, rotational velocity at the equator of 300\,km/s and stellar radius of 20\,R$_\odot$.

In simulations where the core radius is varied, the geometry of the cluster is preserved up to a global rescaling.

In simulations where only 5 stars are considered, only the first 5 rows of Table~\ref{tab:starlist} are implemented, with $\dot{M} = 18 \times 10^{-6}\,\Msun$/yr. All other parameters are kept the same.

\begin{figure}
    \centering
  	\includegraphics[width=0.9\linewidth]{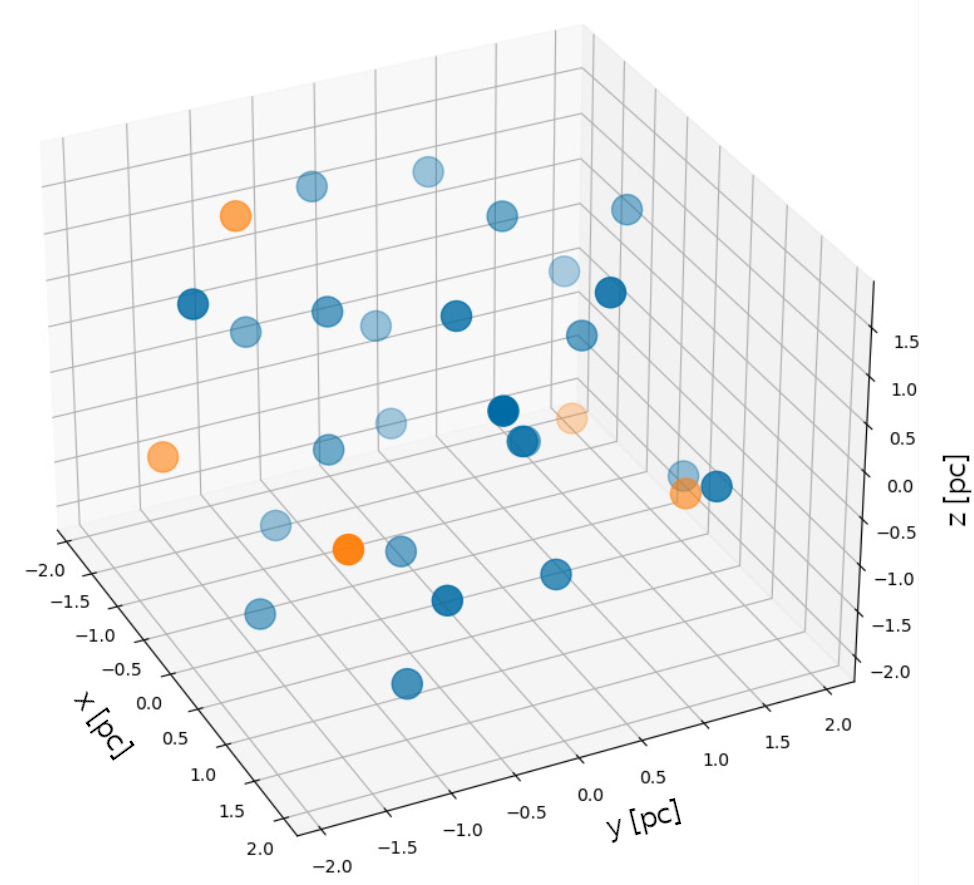}
  	\caption{Distribution of the stars inside the fiducial cluster. Axes are shown in pc. The cluster with 5 stars implements only the orange stars. The opacity of the dots is only a visual cue to visualize the depth.}
  	\label{fig:toycluster}
\end{figure}

Figure~\ref{fig:toycluster} provides a visual impression of the cluster configuration.

\begin{table} 
	\centering
\begin{tabular}{l|l|l|l|l}
$x$ [pc]  & $y$  [pc]  & $z$ [pc]   & $\theta$ & $\phi$   \\
		\hline
		\hline
1.75  & -1.55 & 0.15  & 1.853 & 0.333 \\
-0.90 & -1.20 & 1.70  & 1.776 & 4.449 \\
1.05  & 1.45  & -0.80 & 2.215 & 3.870 \\
-0.95 & 1.55  & -1.40 & 0.355 & 1.981 \\
-0.70 & -1.90 & -0.50 & 2.049 & 3.737 \\
-1.30 & 0.20  & -1.25 & 1.427 & 4.452 \\
0.65  & -0.20 & 1.35  & 1.138 & 4.635 \\
-1.90 & 0.80  & 0.95  & 1.103 & 1.191 \\
-1.35 & -0.95 & 0.10  & 1.446 & 3.262 \\
0.25  & -0.45 & -1.35 & 1.700 & 5.411 \\
1.95  & 1.20  & 0.00  & 0.946 & 1.782 \\
-0.55 & -0.65 & 0.75  & 1.019 & 4.434 \\
-0.15 & -1.40 & -2.00 & 1.180 & 1.231 \\
0.25  & 1.00  & 0.50  & 0.778 & 4.861 \\
-1.40 & 1.70  & -0.10 & 0.629 & 5.800 \\
1.45  & 0.20  & -0.95 & 1.668 & 0.804 \\
-0.65 & 0.80  & 1.25  & 0.805 & 5.116 \\
-0.05 & 0.70  & -0.75 & 2.138 & 3.382 \\
-0.60 & 1.80  & 1.05  & 0.422 & 4.864 \\
1.25  & 0.70  & 1.70  & 0.887 & 5.825 \\
-1.80 & -0.20 & 1.15  & 0.755 & 4.691 \\
1.10  & 0.10  & 0.25  & 1.177 & 5.574 \\
1.75  & -0.80 & -0.65 & 1.938 & 6.042 \\
-0.05 & -1.90 & 1.55  & 2.202 & 5.637 \\
1.90  & -0.45 & 1.30  & 0.597 & 0.718 \\
-1.45 & 0.15  & -0.25 & 0.975 & 0.848 \\
1.30  & -0.90 & -1.85 & 0.939 & 6.024 \\
-1.10 & -0.85 & -1.90 & 1.898 & 1.519 \\
-0.45 & -0.70 & -0.65 & 2.585 & 3.136 \\
0.10  & 1.95  & -1.45 & 2.206 & 4.404
	\end{tabular}
	\caption{Properties of the fiducial cluster with core radius 2.5 pc. $\theta$ and $\phi$ indicate the orientation of the magnetic axis. The cluster with 5 stars implements only the first 5 rows.}
	\label{tab:starlist}
\end{table}

\section{Thermal conduction}
\label{sec:appendix_tc}
\begin{figure*}
    \centering
  	\includegraphics[width=0.75\linewidth]{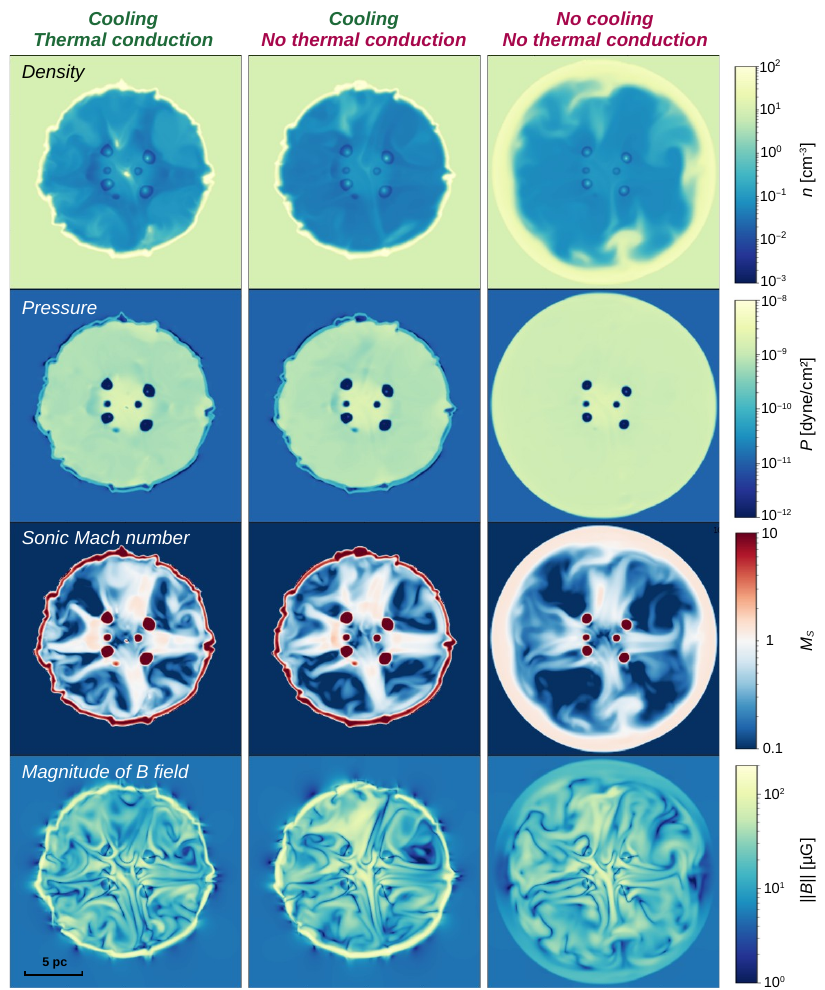}
  	\caption{Comparison of the simulation result in the $(x,y)$ plane at 50\,kyr, showing the influence of thermal conduction and cooling.}
  	\label{fig:Check_thermal_conduction}
\end{figure*}

To test the influence of thermal conduction, we use the standard PLUTO implementation which employs an anisotropic thermal conductivity that preferentially transfers heat along the magnetic field lines. The unsaturated thermal conduction coefficients and description of the saturated regime are given in \citet{Orlando2008,PLUTO2012,badmaev2022}. The parabolic term is integrated with Super-Time-Stepping \citep{PLUTO2007}.

Due to the high resolution of the simulation and high temperatures in subsonic regions, solving for thermal conduction is computationally expensive and we could only reasonably simulate the first few tens of kyr of evolution, starting from the cluster in a homogeneous ISM (same setup as the control simulation described in Section~\ref{sec:HDSimuNascent}).
Figure~\ref{fig:Check_thermal_conduction} shows the result at 50\,kyr. At this time, wind-wind interactions in the core launch two-dimensional transsonic sheets in the superbubble, but the collective cluster outflow has not formed yet.

One can check that including thermal conduction has a noticeable influence. In particular, as expected from analytical theory, heat exchange at the shell interface tends to increase the density in the superbubble. On the other hand, thermal conduction does not impact significantly the overall morphology of the outflows, at least not at these early times.

\section{Test of superbubble ansatz: other slices}
\begin{figure*}
    \centering
  	\includegraphics[width=\linewidth]{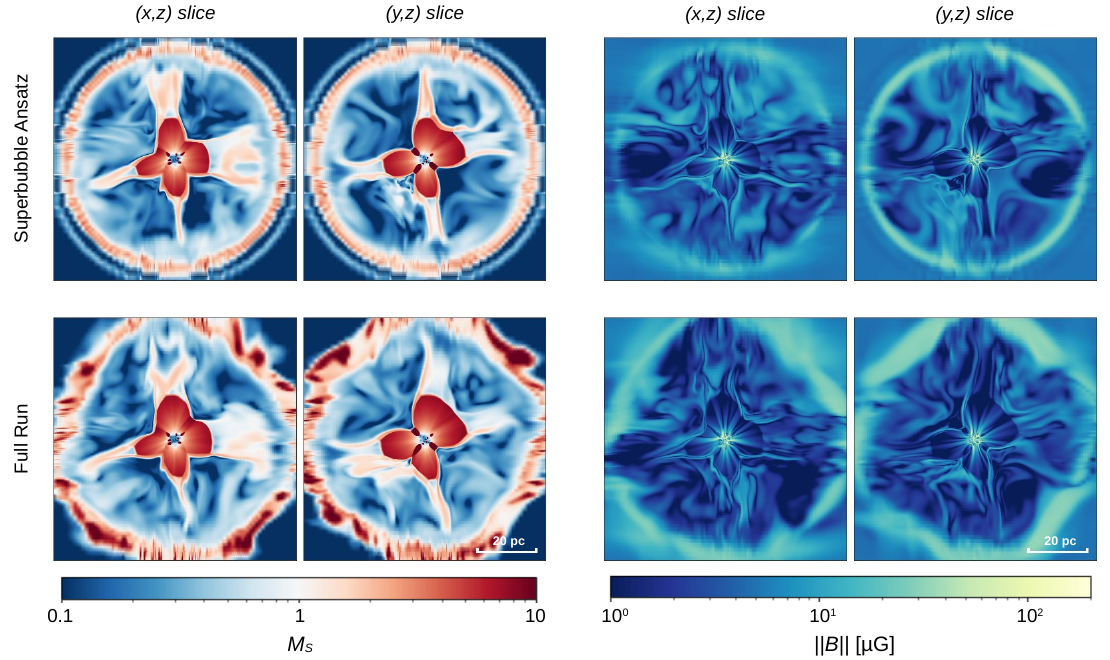}
  	\caption{Sonic Mach number and magnetic field intensity in the $(x,z)$ and $(y,z)$ planes, comparing the result obtained with the superbubble ansatz (top row) and the result obtained in the control simulation (bottom row).}
  	\label{fig:WeaverAnsatzProofat1Myr_otherslices}
\end{figure*}
This appendix complements Section~\ref{sec:proofprinciple}, showing the comparison between the result obtained with the superbubble ansatz and that obtained with the full simulation at 1\,Myr. Figure~\ref{fig:WeaverAnsatzProofat1Myr_otherslices} shows the comparison in two additional slices, for the sonic Mach number and the intensity of the magnitude field. As discussed in Section~\ref{sec:proofprinciple}, the overall good agreement justifies the use of the superbubble ansatz.

\end{appendix}
\end{document}